\documentclass{llncs}

\usepackage{verbatim}
\usepackage{graphicx}
\usepackage{amsmath}
\usepackage{amsfonts}
\usepackage{verbatim}
\usepackage{array}
\usepackage{float}
\usepackage{url}
\usepackage{gensymb}
\usepackage{subfigure}
\makeatletter

\newcommand{\Rmnum}[1]{\expandafter\@slowromancap\romannumeral #1@}
\makeatother

\begin{document}

\title{Evaluation of Zika Vector Control Strategies Using Agent-Based Modeling}
\titlerunning{Mutant Mosquitoes or Bacterial Infection}

\author {Chathika Gunaratne\inst{1,2}%\inst{1, 2}Evolutionary or Contagion Dynamics evolutionary and contagion dynamics based 
\and Mustafa \.{I}lhan Akba\c{s}\inst{1}
\and Ivan Garibay\inst{1, 2}\thanks{Corresponding author: \email{igaribay@ucf.edu}}
\and \\{\"O}zlem {\"O}zmen\inst{1} 
%\and Thomas O'Neal\inst{1, 3}
}
\institute{
Complex Adaptive Systems Laboratory
\and Institute for Simulation and Training \\
University of Central Florida\\
Orlando, Florida, USA \\
}
\maketitle

%ILHAN Abstract is a little bit long and disorganized, we can work on it after results

\begin{abstract} % 150 words
Aedes Aegypti is the vector of several deadly diseases, including Zika. Effective and sustainable vector control measures must be deployed to keep A. aegypti numbers under control. The distribution of A. Aegypti is subject to spatial and climatic constraints. Using agent-based modeling, we model the population dynamics of A. aegypti subjected to the spatial and climatic constraints of a neighborhood in the Key West. Satellite imagery was used to identify vegetation, houses (CO$_{2}$ zones) both critical to the mosquito lifecycle. The model replicates the seasonal fluctuation of adult population sampled through field studies and approximates the population at a high of 986 (95\% CI: [979, 993]) females and 1031 (95\% CI: [1024, 1039]) males in the fall and a low of 316 (95\% CI: [313, 319]) females and 333 (95\% CI: [330, 336]) males during the winter. We then simulate two biological vector control strategies: 1) Wolbachia infection and 2) Release of Insects carrying a Dominant Lethal gene (RIDL). Our results support the probability of sustained Wolbachia infection within the population for two years after the year of release. egies, our approach provides a realistic simulation environment consisting of male and female Aedes aegypti, breeding spots, vegetation and CO$_{2}$ sources.
\end{abstract}

\section{Introduction}
\label{Introduction}

Zika, first identified in Central Africa as a sporadic epidemic disease, has grown into a pandemic with cases reported from every continent within the span of a year. South America is currently most heavily hit with over 4200 suspected cases reported in Brazil itself \cite{Saude2015}. The primary vector of the Zika virus is the Aedes Aegypti mosquito also responsible for the spread of yellow fever, dengue, malaria and chikungunya. 

At the time of writing, the Centers for Disease Control has issued several reports warning the public of the potential devastation of public health that Zika poses in the US. Florida's warm and humid environment, in particular, provides an excellent breeding ground for A. Aegypti. Public health administration departments like the Florida Keys Mosquito Control District have been monitoring and controling mosquito populations in the region. In addition to the traditional population control methods such as destruction of breeding ground through public cleanups, DDT/insecticide spraying, etc. two biological methods have gained popularity in recent years. The first, the Release of Insects carrying a Dominant Lethal gene (RIDL), involves the release of a large number of genetically engineered mosquitoes into the wild \cite{Evans2014}. RIDL uses a `suicidal' gene which prevents the offspring of the genetically modified mosquito from maturing into adulthood. %ILHAN decide whether to use Aedes Aegypti or A. Aegypti
The second method, an incompatible insect technique (IIT), involves the release of mosquitoes infected with the intracellular bacteria, Wolbachia pipentis, which occurs naturally in insects. However at high concentrations, Wolbachia has been proven to reduce the adult lifespan of A. aegypti by up to 50\%\cite{Mcmeniman2009}. 

Both vector control techniques have potential long-term difficulties despite their ability to reduce mosquito numbers upon release. The inability of offspring resulting from RIDL to survive into adulthood also means that the Dominant Lethal gene will not be inherited \cite{Shelly2011}. Therefore, regular releases must be made to maintain long-term sustainability of this approach. On the other hand, Wolbachia infection may transmitted from parent to child through reproduction and remain in the population throughout generations. Yet, spatial and climatic constraints may limit Wolbachia-infected adults from finding mates in the wild or result in infected females being killed off prior to ovipositioning. The production of large volumes of RIDL or Wolbachia-infected A. aegypti may be costly. Attempts to establish a sustained infection of Wolbachia within A. aegypti populations in the wild have been attempted \cite{Nguyen2015}. Therefore, identifying the long-term sustainability and required release volumes of mosquitoes is important.

Despite the difficulty of suppressing the mosquito population as a whole, A. aegypti is quite vulnerable to climatic and spatial conditions on an individual scale. In particular, the fetal/aquatic lifespan (time spent in egg, larval and pupal stages), adult lifespan, mortality rates and probability of emergence are highly sensitive to variations in the temperature. The Key West, despite having a tropical climate with a yearly average temperature range of 10 \degree C , has been shown to have a reasonable fluctuation in mosquito population throughout the year. 

In addition to climatic variations, mosquito survival is heavily dependent on abundance of vegetation, human hosts and breeding sites. The male mosquito depends on vegetation for food, while it is the female mosquito that feeds on the blood of mammals. The female mosquito is attracted to hosts by CO\_2 and pheromone emissions and can detect hosts from upto 30m away \cite{Cummins2012,Gillies1972}. Vegetation zones must be within reasonable proximity of host locations in order for males to be able to reach females for mating. Finally, there must be an abundance of breeding sites (exposed stagnant water) upon which females must oviposition (lay eggs).

In an effort to identify the sustainability of the two vector control techniques, we use agent-based modeling to simulate the yearly fluctuation of mosquito population dynamics in the Key West. A suburban neighborhood is selected and segmented into vegetation, houses($CO_2$ zones) and breeding zones to capture the spatial constraints experienced by the local mosquito population. Satellite imagery of the neighborhood is processed to identify the exact location of these zones. In addition to the spatial constraints, the monthly temperature variation of the Key West is also simulated as a climatic constraint. Mosquito agents are released into this environment and their population characteristics are observed throughout time. After validating the yearly adult population fluctuation produced by this model, we use it to simulate and compare the two vector control strategies mentioned.

\section{Background}
\label{RelatedWork}

Modeling and simulation have been used to study environmental and animal monitoring problems \cite{Arifin2015} \cite{Brust2013} \cite{Akbas2010}. For the mosquito population dynamics modeling, there is a variety of approaches in the literature including analytical models, differential equation models and ABMs. One of the more prominent mosquito dynamics models in the literature is CIMSim \cite{Focks1993}, uses dynamic life-table modeling of life-stage durations of the aedes gonotrophic cycle, as influenced by environmental conditions such as temperature and humidity. Despite its lack of spatial properties, CIMSim has been recognized as the standard mosquito population dynamics model by the UNFCCC (United Nations Framework Convention on Climate Change). Other similar models include DyMSim \cite{Morin2010}, TAENI2 \cite{Ritchie1995} and the use of Markov chain modeling in \cite{Otero2006}. A spatially explicit version of CIMSim, Skeeter Buster is also commonly used for mosquito population estimation \cite{Magori2009}. 

ABMs differ from the other models by capturing the spatial interactions among individuals which emerge into macro scale results of small changes in individual characteristics or behaviour of the agents. Our approach employs a spatial model of the A. aegypti population by integrating an ABM with geographic information. Spatial models are used in epidemiology to study population dynamics or to evaluate methods for population control. Evans and Bishop \cite{Evans2014} propose a spatial model based on cellular automata to simulate pulsed releases and observe the effects of different mosquito release strategies in Aedes aegypti population control. The results of the model show the importance of release pulse frequency, number of release sites and the threshold values for release volume. 

Another spatial approach for simulating Aedes aegypti population is SimPopMosq \cite{Almeida2010}, an ABM of representative agents for mosquitoes, some mammals and objects found in urban environments. SimPopMosq is used to study the active traps as a population control strategy and includes no sterile insect agents or techniques. The framework by Arifin et al. \cite{Arifin2015} integrates an ABM with a geographic information system (GIS) to provide a spatial system for exploring epidemiological landscape changes (distribution of aquatic breeding sites and households) and their effect on spatial mosquito population distribution. Lee et al. \cite{Lee2013} also investigate the influence of spatial factors such as the release region size on population control. The method uses a mathematical model to study the relation among the location related parameters. Isidoro et al. \cite{Isidoro2009} used LAIS framework to evaluate the RIDL for Aedes aegypti population. The ABM includes independent decision-making agents for mosquitoes and pre-determined rule based elements for environmental objects such as oviposition spots. However the model lacks important factors such as a realistic map or temperature effects. An observation in most of these studies is the lack incorporation of male mosquito dynamics and their requirement to travel between vegetation for nutrition and mates. 

There are also approaches integrating the mosquito population control models with epidemic models. Deng et al. \cite{Deng2008} proposed an ABM to simulate the spread of dengue, the main vector of which is Aedes aegypti as well. The mobility of the mosquitoes in this model are defined by a utility function, which is affected by the population, wind and landscape features. However, the model lacks a granular spatial discretization and only a small number of agents are used. Moulay and Pign\'{e} \cite{Moulay2013} studied Chikungunya epidemic with a metapopulation network model representing both mosquito and human dynamics on an island. The model is created by considering both the density and mobility of populations and their effects on the transmission of the disease.

\section{Methodology}
\label{OurProtocol}

We model the population dynamics of mosquitoes in an agent-based model implemented in RePast Simphony \cite{North2013} consisting of agents embodying the behavior of A. aegypti and feeding and breeding off of designated zones in a geographical environment with monthly changes in average temperature. The distribution of the zones provided spatial constraints on the total population while changing temperature applied climatic constraints. Zones were either locations with CO$_2$ (human hosts), vegetation or breeding sites. The distribution of these zones were determined using geographical analysis of a suburban neighborhood in Key West, Florida. The monthly average temperature in Key West was obtained from \cite{US2016}.

\subsection{Life Stages, Processes, Circadian Rhythm and Behavior Modes}
A. aegypti has four life stages and undergoes metamorphosis between these stages. The first three life stages (egg, larva and pupa) are spent in water while the final stage (adult) is spent as an airborne insect. Adult females feed on blood of mammal hosts, while males gain nutrition from vegetation. Female mosquitoes are attracted to hosts through $CO_2$ and pheromones upon which the perform a process known as klinotaxis to reach their host. The female A. aegypti prefers to lay eggs closer to urban areas and is considered a domestic pest. 

\begin{figure*}[ht]
\begin{center}       
	\includegraphics[width=0.8\textwidth]{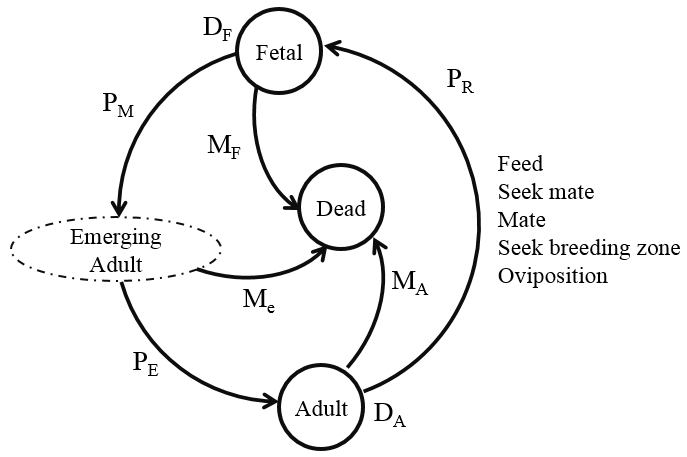}		
	\caption{Lifecycle of the mosquito agent. }
	\label{Fig:LifeCycle}
\end{center}
\end{figure*}

The life stages of A. aegypti are simulated in our model. The lifecycle of the simulated mosquito agents is described in Fig. \ref{Fig:LifeCycle}. For the purpose of our study, the egg, larva, and pupa stages were considered as a single stage, FETAL, and considered to be inanimate. During the FETAL stage the mosquito remains within the confines of the breeding site. A FETAL has a probability of dying $M_F$ (mortality rate, probability of maturing: $P_M = 1 - M_F$). Once the FETAL agent has survived for $D_f$ days, it emerges into an adult. Emergence is probabilistic and there is $M_E$ chance of death during emergence (probability of emergence: $P_E = 1 - M_E$).

Emerged ADULTs live for $D_A$ days. ADULTs may die during their life processes or due to old age at a rate of $M_A$. New FETALs are created through reproduction with a probability of $P_R$. $P_R$ depends on an individual adults ability to find food sources, feed, seek mates, mate successfully, seek breeding zones and oviposition. These processes are constrained by the spatial distribution of the zones and restriction of $D_A$ due to temperature. Further, mating success is probabilistic (probability of successful mating: $p_m$). Adult females may be killed by human hosts while feeding (daily probability of female being killed by human host: $p_h$). $M_A$ and $P_R$, are therefore, subject to various factors and highly variable depending on the individual mosquito's sex, location in relation to other mosquitoes, location in relation to zones and the temperature of the environment. However, precalculation of $M_A$ and $P_R$ are not required due to the computational nature of agent-based modeling. 

\begin{figure*}[ht]
\begin{center}       
	\includegraphics[width=0.8\textwidth]{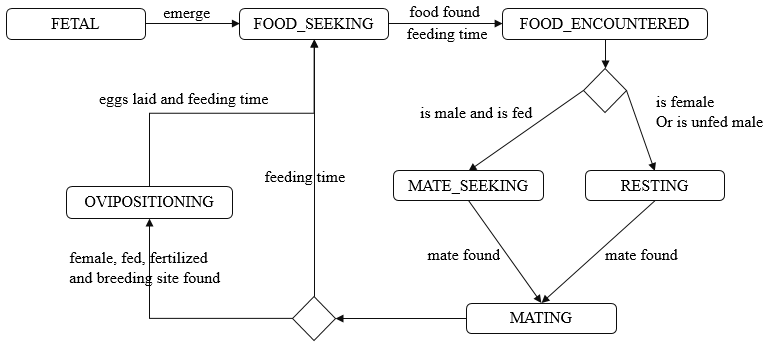}		
	\caption{State diagram for the adult mosquito agent. }
	\label{Fig:StateDiagram}
\end{center}
\end{figure*}

In our model, all adult mosquitoes emerge from the FETAL process into the FOOD\_SEEKING process. As shown in Figure \ref{Fig:StateDiagram}, when in range of an appropriate food source, the agent switches to the FOOD\_ENCOUNTERED process. The female mosquito searches for blood meals by seeking out CO$_{2}$ sources within the environment, while males seek out vegetation zones. After a period of feeding, the mosquitoes enter the mating phases. The female mosquito agents transition to RESTING until fertilization, upon which she enters into the OVIPOSITIONING phase. Meanwhile, male mosquitoes enter the MATING phase and seek out potential mates, until their energy is depleted upon which they enter the RESTING phase. This completes the daily rhythm of the mosquito.

There are certain conditions of satisfaction for the mosquito agents to transition from one process to another as described in Figure \ref{Fig:StateDiagram}. In order for a mosquito to enter into any of the processes described above other than the FETAL process, it must be in the ADULT phase. In order for a female to produce eggs, it must have enough energy or be fed. To enter OVIPOSITIONING, the female must also be fertilized by a male mate.

The adult mosquito agents in the model follow a daily rhythmic behavior depending on their current state. A. aegypti circadian rhythms reported by Chadee \cite{Chadee2013} demonstrated that blood feeding, oviposition, sugar feeding and copulation occurred mostly between 06-09 hours and between 16-18 hours. The mosquitoes rest for the remaining time of the day except atypical biting. Hence, the daily time was partitioned into eight equal segments in our model. Following the information given by Chadee \cite{Chadee2013}, ovipositioning was allowed during the second and fifth segments of the day while feeding was allowed during the second, third, fifth and sixth segments of the day.

Climatic constraints on mosquitoes are considered to occur through varying monthly temperature. Field studies of A. aegypti in the wild and laboratory experiments have established the relationships of average temperature and mortality rates, probability of adult emergence from pupa and life stage durations. There are several studies in the literature \cite{Yang2009,Brady2013}, which have fit mathematical models relating aquatic/fetal mortality, adult mortality, fetal duration, adult duration and probability of emergence with temperature. Accordingly the model allows for temperature dependency of several parameters effecting the mosquito lifecycle including FETAL and ADULT mortality rates and durations and oviposition rate.

\subsection{Geographical Environment}
\begin{figure*}[ht]
\begin{center}       
	\includegraphics[width=0.8\textwidth]{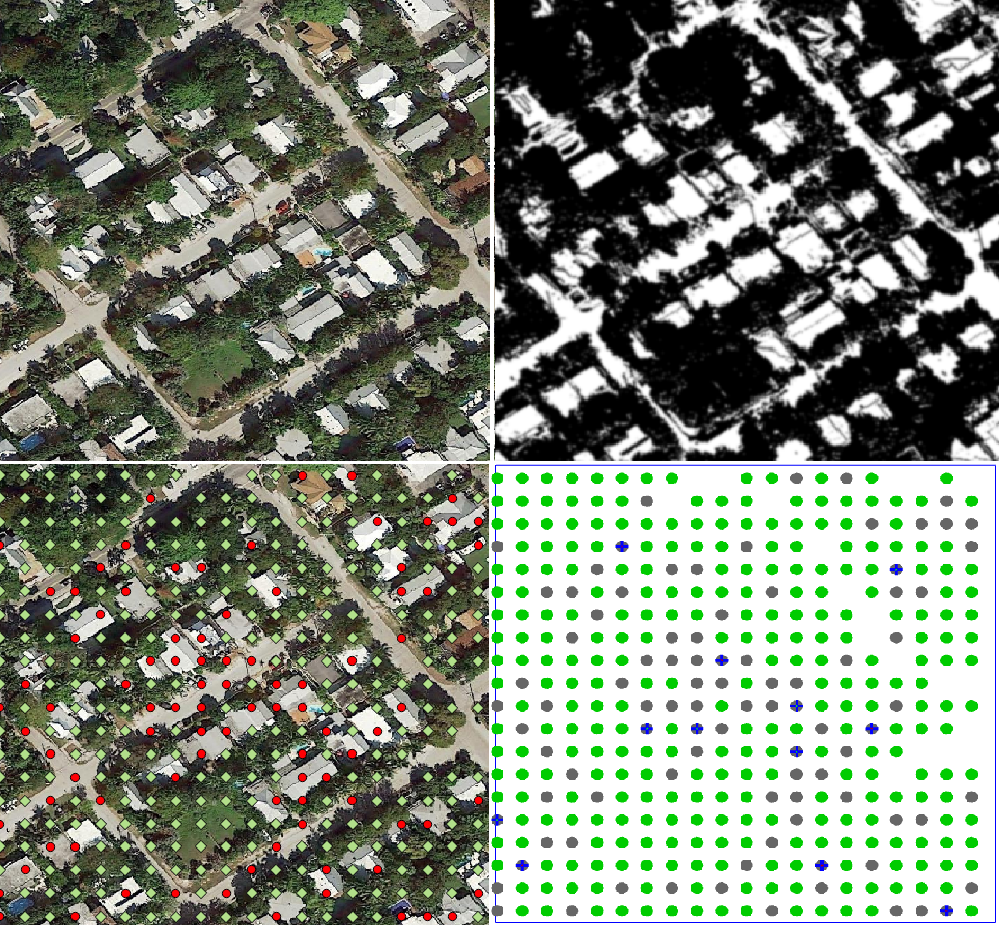}		
	\caption{Satellite imagery of the suburban neighborhood simulated in the study being processed and converted to zones simulation in RePast.}
	\label{Fig:ProcessedImage}
\end{center}
\end{figure*}

The simulations were run on a suburban neighborhood (Lat: -81.78095, Lon: 24.55350) in the Key West, FL. An area of 29584 $m^2$ was simulated consisting of two blocks of housing. Satellite imagery was obtained through Google Earth and processed using QGIS (Fig. \ref{Fig:ProcessedImage}(top-left)). After geomapping of the satellite image and noise cancelation, the image was converted to grayscale and segmented through a k-means unsupervised learning algorithm searching for two classes by pixel intensity (Fig. \ref{Fig:ProcessedImage}(top-right)). The resulting polygons were then overlain with a regular grid of points. Each point having 10m spacing between them. The points were then classified according to which class of polygon they intersected on the map image. The result was a representation of the distribution of vegetation zones and urban areas in this neighborhood  (Fig. \ref{Fig:ProcessedImage}(bottom-left)). 

The point layer was then imported into Repast as an ESRI shapefile. Each point was then made the center of a circular vegetation zone or CO$_{2}$ source with radius ($R_C$) or ($R_V$), respectively. 

The prevalence of breeding zones depended on the house index (breeding sites per house per week) in the region. The average house index as reported by FKMCD was approximately 20\% in 2010\cite{FKMCD2013a,FKMCD2013b}. Accordingly, 20\% of the CO$_{2}$ zones were, randomly, also designated as breeding zones with radius ($R_B$). An example of the distribution of zones within the simulated region is shown in (Fig. \ref{Fig:ProcessedImage})(bottom-right).

\subsection{Vector Control Strategies}
Superinfection of mosquito populations in the wild with the naturally occurring intracellular bacteria, Wolbachia (also referred to as Incompatible Insect Technique (IIT)) result in Cytoplasmic Incompatibility. Crosses between infected males and uninfected females result in no offspring and has been used in suppression of mosquito populations in the wild \cite{Zabalou2009}. Most pathogens transmitted by mosquitoes require a development period before they can be transmitted to a human host \cite{Mcmeniman2009}. The time period from pathogen ingestion to potential infectivity, the extrinsic incubation period (EIP), is about 10 days for Zika. Wolbachia has been shown to reduce the lifespan of A. aegypti by upto 50\% \cite{Mcmeniman2009}. Reduced life time of adult female mosquitoes leads to a reduction in the probability of adult female mosquitoes biting humans and resultantly mitigates the transmission of vector-borne disease such as Zika. Sustained Wolbachia infection has been induced in wild mosquito populations by releasing infected females (crosses between infected females and uninfected or infected males results in Wolbachia infected offspring) \cite{Joubert2016,Hoffmann2014,Mcmeniman2009}.

On the other hand, RIDL depends on the artificial genetic alteration of the mosquito to become dependent on tetracycline. Mosquitoes reared in the laboratory are provided on tetracycline and then released into the wild. The resulting offspring die before reaching adulthood due to the absence of tetracycline in the wild. RIDL mosquitoes are usually male, to avoid increasing human-biting mosquitoes by releasing females \cite{Harris2011}. Further, unlike Wolbachia infection, female release is unnecessary since a sustained introduction of RIDL cannot be maintained as all offspring are killed. Potential disadvantages of RIDL have been discussed in \cite{Shelly2011} 

Mosquito agents in the model could be infected with Wolbachia. Mating between uninfected females and infected males results in $D_A = 0$ for all offspring. Mating between infected females and uninfected/infected males results in all offspring being infected with Wolbachia. $D_A$ of these offspring will be halved. 

Mosquito agents may carry the RIDL gene. Only released RIDL mosquitoes will be able to survive in the environment as adults. All children resulting from a RIDL parent will inherit RIDL and set $D_A = 0$. Finally, for RIDLs $p_m = 0.5$ as reported in \cite{Shelly2011}.

\section{Experiments}
\label{SimulationStudy}
The agent-based model was used to estimate the Aedes aegypti population in the Key West. The monthly population flucuation matched that shown in catch rates from the FKMCD \cite{FKMCD2013a,FKMCD2013b}. Populations were lowest during late winter and highest during the summer and late Fall months. 
The model was then used to evaluate the two control strategies (RIDL and Wolbachia infection) over a period of three years. For each experiment, the simulation was allowed to run for 2 simulation years prior to data collection, in order to allow the agents to fit the constraint patterns of the environment. Data collection was performed after the 2$^{nd}$ simulation year and performed for 3 simulation years. 
FKMCD \cite{FKMCD2013a,FKMCD2013b} findings indicate the mean maximum of Aedes Aegypti caught in traps set up near households is 20 per trap per night. Hence, our simulations were initialized with 20 larvae in each breeding site. Values of the other parameters used in all simulation experiments and their sources are listed in table \ref{Tab:Params}. 

\begin{table}
\begin{center}      
\caption{Parameters used in the model ($T$ : Monthly Temperature)}
\begin{tabular}{clcr}
\hline\noalign{\smallskip}
$Parameter$ & $Definition$ & $Value$ & $Source$\\
\noalign{\smallskip}
\hline
\noalign{\smallskip}
$spd$ & displacement speed & 0.5 - 1 m/s & \cite{Almeida2010,Cummins2012} \\
$D_{f1}$ & Mean duration of egg stage & $f(T)$ & \cite{Otero2008} \\
$D_{f2}$ & Mean duration of larval and pupal stages & $f(T)$  & \cite{Yang2009} \\
$D_F$ & Mean duration of FETAL stage & $D_{f1} + D_{f2}$& \- \\
$D_A$ & Mean duration of ADULT stage & $f(T)$  & \cite{Yang2009} \\
$M_F$ & FETAL mortality rate & 0.3 & \cite{Yang2009} \\
$m_l$ & Probability of successful emergence & 0.3 & \cite{Yang2009} \\
$r_c$ & Detection range for CO$_{2}$ zones & 30 m & \cite{Almeida2010,Cummins2012} \\
$r_v$ & Detection range for vegetation zones & 30 m & \cite{Almeida2010,Cummins2012} \\
$r_v$ & Detection range for breeding zones & 30 m & \cite{Almeida2010,Cummins2012} \\
$r_m$ & Detection range for mates & 30 m & \cite{Almeida2010,Cummins2012} \\
$r_m$ & Number of mates per male per day & 5 & \cite{Choochote2001} \\
$r_m$ & Probability of successful mating & 0.7 & \cite{Almeida2010} \\
$r_m$ & Number of times a female can lay eggs in one lifetime & 5 & \cite{Choochote2001} \\
$r_m$ & Eggs laid in one oviposition & 63 & \cite{Evans2014} \\
$r_m$ & Duration of one oviposition & 3-4 days & \cite{Evans2014} \\
$d_w$ & ADULT duration decrease due to Wolbachia & 50\% & \cite{Mcmeniman2009} \\
$d_l$ & ADULT duration decrease due to lethal gene & 100\% & \cite{Shelly2011} \\
$d_l$ & Mating success of RIDL males & 50\% & \cite{Shelly2011} \\
\hline
\end{tabular}
\label{Tab:Params}
\end{center}      
\end{table}

\subsection{Population Estimation}

\begin{figure}[!hp]
\centering
     \subfigure{\includegraphics[width=\textwidth]{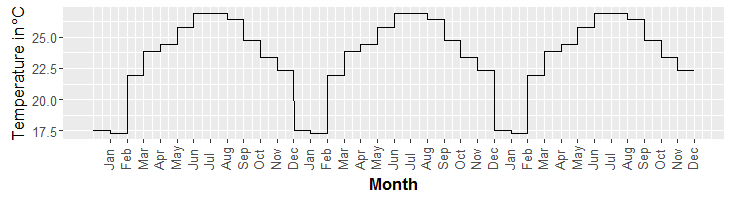}}
     \caption{Monthly temperature fluctuation over three years}
     \label{Fig:Temperature}
\end{figure}
\begin{figure}[!hp]
\centering
     \subfigure{\includegraphics[width=\textwidth]{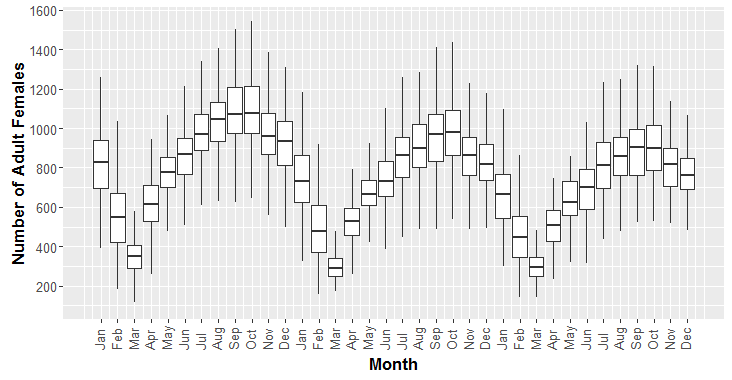}}
     \caption{Simulation results for female adult population over three years}
     \label{Fig:Pop_Est_F}
\end{figure}
\begin{figure}[!hp]
\centering
     \subfigure{\includegraphics[width=\textwidth]{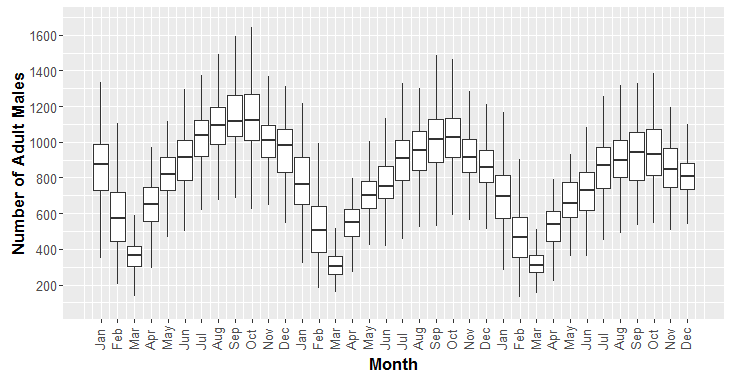}}
     \caption{Simulation results for male adult population over three years}
     \label{Fig:Pop_Est_M}
\end{figure}
Using the model described above we were able to make estimations on the Aedes aegypti population in the Key West neighborhood considered. It was seen that the adult populations closely followed the monthly temperature fluctuations (Figure \ref{Fig:Temperature}). As shown in figures \ref{Fig:Pop_Est_F} and \ref{Fig:Pop_Est_M}, adult populations were highest during October and lowest during March. The male population slightly exceeded the female population. During October, the mean count of females was 986 (95\% CI: [979, 993]) and the mean count of males was 1031 (95\% CI: [1024, 1039]). In March, the mean count of females was 316 (95\% CI: [313, 319]) and the mean count of males was 333 (95\% CI: [330, 336]).
\subsection{Simulating Vector Control Strategies}
We adopted vector control strategies from field trials for both the Wolbachia technique and RIDL. Attempts have been made to establish a sustained Wolbachia infection in the Aedes aegypti population in Machans Beach, Australia \cite{Nguyen2015}. We simulated the same release quantities per urban zone in our model on the Key West, to reflect the release quantities used in the field trial. This resulted in two releases being simulating. The first release consisted of 253 males and females each, weekly, over a period of 15 weeks. In the second release 138 males were released weekly, over a period of 10 weeks. Releases were performed at every fourth urban zone (as in the field trials) and initiated in the first week of April. A total of 8970 adults were released. 

The release strategy for for RIDL was adopted from field trials conducted in the Cayman Islands \cite{Harris2011}. To allow for comparison a total of 8970 adults were released over 25 weeks in each simulation run. 368 males were released over the first 24 weeks and 138 in the last week. Again releases were performed at every fourth urban zone. 

For both the Wolbachia and RIDL cases, data was aggregated over 90 runs. As release periods for both cases was 25 weeks, the final release was on the first week of October in both cases. 
\begin{figure*}[hp]
\begin{center}       
	\includegraphics[width=\textwidth]{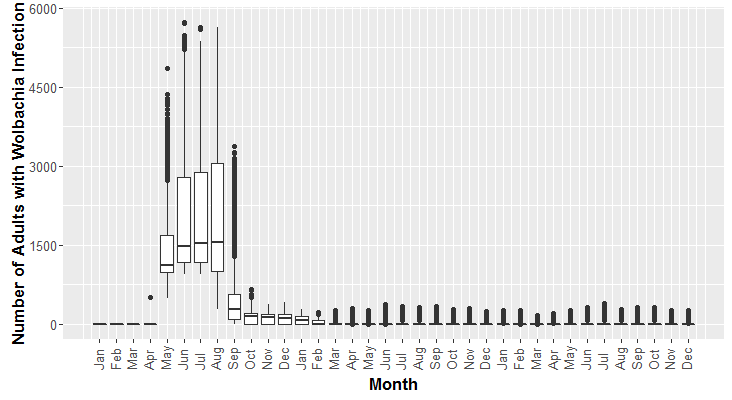}		
	\caption{Simulation results for the number of adults with Wolbachia infection over three years}
	\label{Fig:WolSusAgg}
\end{center}
\end{figure*}
\begin{figure*}[hp]
\begin{center}       
	\includegraphics[width=\textwidth]{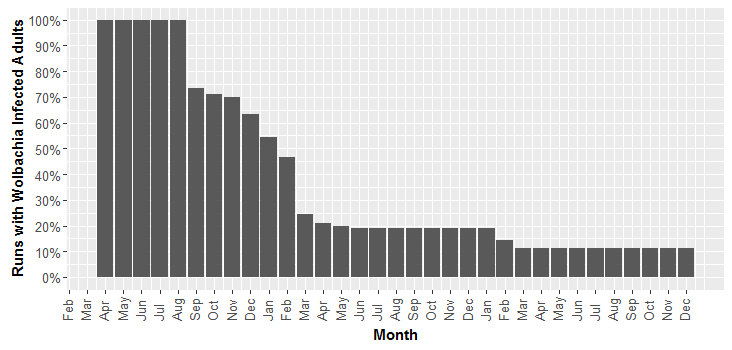}		
	\caption{Percentage of simulation runs which sustained wolbachia infection over the three year period}
	\label{Fig:WolSus}
\end{center}
\end{figure*}
For the purpose of this study, we observed the prevalence of Wolbachia infected adults and adults carrying the dominant lethal gene, during and after the release period, for each case. Figure \ref{Fig:WolSusAgg} demonstrates the aggregate abundance of Wolbachia infected adults within the population. It can be seen that Wolbachia infection remained within the population even when the total mosquito population dropped during the colder months. As seen in figure \ref{Fig:WolSus} around 11\% of the runs did manage to sustain Wolbachia infection within the population for 2 years after the release period.
\begin{figure*}[h]
\begin{center}       
	\includegraphics[width=\textwidth]{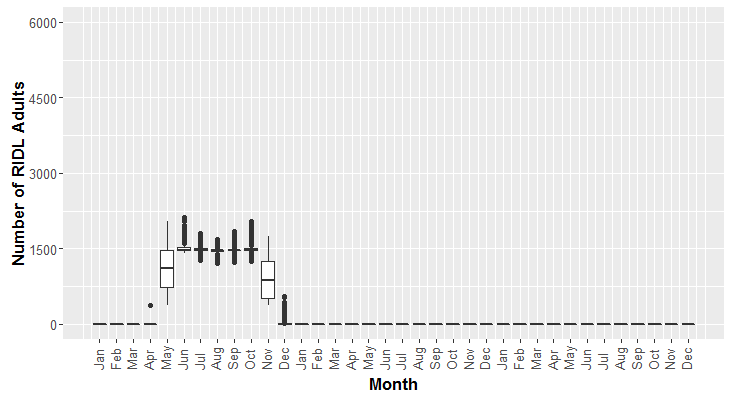}		
	\caption{Simulation results for the number of adults carrying the dominant lethal gene over three years}
	\label{Fig:RIDLSusAgg}
\end{center}
\end{figure*}
As seen in figure \ref{Fig:RIDLSusAgg}, the number of mosquitoes carrying the dominant lethal gene dropped back to zero as soon as releases were discontinued and the released generation had died out. 

\section{Conclusion}
\label{Conclusion}
We have designed an agent-based model of the mosquito population in the Key West, Florida in an effort to address the control of the Zika pandemic. The primary vector of Zika, Aedes aegypti was modeled on a geographical space representing a suburban neighborhood. Satellite imagery was used to capture the spatial distribution of households ($CO_2$ zones), vegetation zones and breeding sites. Additionally, the monthly variation in temperature in the Key West was simulated. Using these spatial and climatic constraints the annual cycle of the mosquito population was replicated by the model to match trends demonstrated by weekly catch rates reported in field studies. It was shown that the spatial and climatic constraints in the Key West allowed for a maximum of approximately 986 (95\% CI: [979, 993]) females and 1031 (95\% CI: [1024, 1039])females in the late Fall, while in the late winter the population remained at a low of 316 (95\% CI: [313, 319]) females and 333 (95\% CI: [330, 336]) males.

Two vector control strategies were simulated using the described ABM. The first strategy, the release of Wolbachia infected mosquitoes, involved releasing male and female mosquitoes with high levels of Wolbachia infection. The release strategy, including release quantities, ratios and frequency, followed a field trial performed in Machans Beach, Australia \cite{Nguyen2015}. Infected males that mated with uninfected females would result in dead offspring, while infected females would produced offspring with Wolbachia infection.

The second strategy, Release of Insects carrying a Dominant Lethal gene (RIDL), involved releasing males that would produce offspring that could not survive into adulthood. If these males competed successfully with wild males for mates, then the population would reduce as a result. The RIDL release strategy followed a field trial performed in the Cayman Islands \cite{Harris2011}. The total volume of RIDL males released was equal to the total Wolbachia infected mosquitoes released to allow for comparison. 

It was observed that Wolbachia infection could be established within a population of Aedes aegypti in the Key West. However, the low probability of establishing sustained infection (approximately 11\%) suggested that infection was highly susceptible to uncertainty of the environment. One of the major factors of uncertainty in the model was the spatial orientation of the breeding sites. Therefore, there is evidence to believe that the spatial orientation of the breeding sites has an impact on where releases must be performed in order to maintain Wolbachia infection within the population. Similar observations have been made in the field \cite{Nguyen2015}, however, further analysis must be performed in order to confirm this conclusion. 

Contrastingly, the model also demonstrated the inability of the RIDL technique to be established within the population. This result is expected as the dominant lethal gene is not inherited into future generations due to the death of all progeny of the released mosquitoes. 

Finally, we have shown that this model can be used to simulate what-if scenarios and experiment with the release volumes and frequencies of vector control strategies for A. aegypti. The spatial and climatic constraints captured in this model allow it to closely represent the distribution of A. aegypti in Key West and the same technique can be applied for any geographical location. 
%\balance
\bibliographystyle{ieeetran}
\bibliography{Evaluation_of_Zika_Vector_Control_Strategies_Using_Agent-Based_Modeling}

\end{document}